\documentclass[12pt]{article}
\usepackage{amssymb}
\def\pacs{\rightline}

\begin{document}

\date{\today}

\begin{center}
{\large \textbf{QUANTUM COMPUTER AND ITS QUASICLASSICAL MODEL}}

{\medskip TIMUR F.~KAMALOV }

{MEGALLAN ltd., pr. Mira,180-3,Moscow129366,Russia\\[0pt]
Tel./fax 7-095-282-14-44\\[0pt]
E-mail: okamalov@chat.ru \& ykamalov@rambler.ru }
\end{center}

Could the theories with hidden variables be employed for creation of a
quantum computer? A particular scheme of quasiclassical model quantum
computer structure is describe.

\bigskip

In quantum theories with hidden variables the wavefunction $\Psi =\Psi
\left( u_{i}\right) $ being the functions of hidden variables. Let us
consider the following mental experiment. Let us connect two classical
computers by the quantum cell, the latter being the embodiment of the EPR
experiment~\cite{Einstein}. For this purpose, let us consider the fission of
a zero-spin particle into two particles with non--zero spins. Wavefunction
of these are $\Psi _{A}=\Psi \left( q_{kA},S_{A},u_{i}\right) $ and $\Psi
_{B}=\Psi \left( q_{kB},S_{B},u_{i}\right) ,$ $q_{i}$ being the coordinate, $%
S_{A},$ $S_{B}$~--- the spins of $A$ and $B,$ $u_{i}$~--- hidden variables.
Having processed the information at computers $A$ and $B,$ we shall obtain
at the third classical computer the interference pattern of $\Psi _{A}$ and $%
\Psi _{B}$ mapped by $\Psi _{AB}=\Psi \left(
q_{kA},q_{kB},S_{A},S_{B},u_{i}\right) .$ Employing this interference, the
quantum computer would enable processing of information of useful signal
with sub--noise level. To implement this scheme, two--photon radiation can
be employed; instead of the latter, multi--photon radiation could be used.

A modern classical computer comprises semiconductor classic bits with two
Boolean states ``0'' and ``1''; these could be, for example, two distinct
values of electric current or potential at a given bit. In a quantum
computer, the basis is a qubit (quantum bit), the wavefunction of which for
the basis states $\left| 0\right\rangle $ and $\left| 1\right\rangle $ is a
superposition $\left| \Psi \right\rangle =\alpha \left| 0\right\rangle
+\beta \left| 1\right\rangle ,$ $\alpha $ and $\beta $ being the complex
amplitudes of state (with $\left| \alpha \right| ^{2}+\left| \beta \right|
^{2}=1$) with probabilities ${P\left( 0\right) =\left| \alpha \right| ^{2},}$
$P\left( 1\right) =\left| \beta \right| ^{2}.$ With rotation of the state
vector $\left| \Psi \right\rangle $ in Hilbert two--dimensional state space,
amplitudes $\alpha $ and $\beta $ vary. If there exists a register
comprising $\lambda $ qubits, then in quantum calculation the unitary
operation of all $2^{\lambda }$ amplitudes is performed. With this, all
qubit must be correlated.

A quantum computer on correlated photons could be arranged as follows. The
qubit of such a quantum computer must be correlated photons. It should be
noted that in case of multiple reflections of a photon, its amplitude is
reduced, its magnitude being dependent on reflection coefficient. Correlated
photons could be arranged according to the EPR experiment scheme, in which
usually two correlated photons are produced. Prior to input information into
the quantum computer, all qubit of the register must be initialized, i.~e.
brought to the main basis states $\left| 0_{1}\right\rangle ,\left|
0_{2}\right\rangle ,\ldots \left| 0_{\lambda }\right\rangle .$ This could be
easily done by an information--input polarizer.

The scheme of correlated--photon quantum computer shall have the form:

\[
\begin{array}{cc}
& \nu _{1}\rightarrow P_{1}^{^{\prime }}\rightarrow \left|
0_{1}\right\rangle \rightarrow P_{1}^{^{\prime \prime }}\rightarrow
U_{1}\rightarrow \left| \Psi _{1}\right\rangle \rightarrow P_{1}^{^{\prime
\prime \prime }}\rightarrow D_{1}\medskip \\ 
& \nu _{2}\rightarrow P_{2}^{^{\prime }}\rightarrow \left|
0_{2}\right\rangle \rightarrow P_{2}^{^{\prime \prime }}\rightarrow
U_{2}\rightarrow \left| \Psi _{2}\right\rangle \rightarrow P_{2}^{^{\prime
\prime \prime }}\rightarrow D_{2}\medskip \\ 
A\phantom{\cdots} & \cdots \cdots \cdots \cdots \cdots \cdots \cdots \cdots
\cdots \cdots \cdots \cdots \cdots \cdots \cdots \medskip \\ 
& \nu _{\lambda }\rightarrow P_{\lambda }^{^{\prime }}\rightarrow \left|
0_{\lambda }\right\rangle \rightarrow P_{\lambda }^{^{\prime \prime
}}\rightarrow U_{\lambda }\rightarrow \left| \Psi _{\lambda }\right\rangle
\rightarrow P_{\lambda }^{^{\prime \prime \prime }}\rightarrow D_{\lambda
}\medskip
\end{array}
\]

\noindent where: $A$ is the point of production of photons, $P_{i}^{^{\prime
}},P_{i}^{^{\prime \prime }},P_{i}^{^{\prime \prime \prime }}$ are
polarizer, $U_{i}$ is rotation of the state vector, $D_{i}$ are detectors.

Let us illustrate the above with a physical model with gravity background
(i.~e. the background of gravity fields and waves) illustrating the 
hidden variables [4-9].

\bigskip Relative oscillations $\ell ^{i}$, $i=0,1,2,3$ of two particles in
gravity fields are described by deviation equations

\begin{center}
\bigskip $\frac{D^{2}}{D\tau ^{2}}\ell ^{i}=R_{kmn}^{i}\ell ^{m}\frac{dx^{k}%
}{d\tau }\frac{dx^{n}}{d\tau }$,
\end{center}

where $R_{kmn}^{i}$ - is the Rieman's tensor.

In this particular case the deviation equations are converted into
oscillation equations for two particles: 
\[
\stackrel{..}{\ell }^{1}+c^{2}R_{010}^{1}\ell ^{1}=0,\quad \omega =c\sqrt{%
R_{010}^{1}}. 
\]

It should be noted that relative oscillations of micro objects $P$ and $Q$
do not depend on the masses of these, but rather on Riemann tensor of the
gravity field. This is important, as in the microcosm we deal with small
masses. Taking into consideration the gravity background, the micro objects $%
A$ and $B$ shall be correlated. It is essential that in compliance with the
gravity theory the deviation equation does only make sense for two objects,
and it is senseless to consider a single object. Therefore, the gravity
background complements the quantum--mechanical description and plays the
role of hidden variables. On the other hand, the von Neumann theorem on
impossibility of hidden variables introduction into quantum mechanics is not
applicable for pairwise commuting quantities (Gudder's theorem~\cite{Gudder}%
). Introduction of hidden variables in the space with pairwise commuting
operators is appropriate.

The solution of the above equation has the form $\ell ^{1}=\ell _{0}\exp
\left( k^{a}x_{a}+i\omega t\right) ,$ $a=1,2,3.$ Here we assume the gravity
background to have a random nature and should be described, similarly to
quantum--mechanical quantities, with variates. Each gravity field or wave
with the index $i$ and Riemann tensor $R\left( i\right) $ should be matched
by a quantity $\ell \left( i\right) $ square of which is the probability of
the particle being located in the given point. It should be noted that the
definition of $\ell \left( i\right) $ is similar to the definition of a
quantum--mechanical wavefunction $\Psi \left( i\right) .$ Therefore, taking
into account the gravity background, i.~e. the background of gravity field
and waves, the particles take on wave properties described by $\ell \left(
i\right) .$

We will only consider in the present study the gravity fields and waves
which are so small that alter the variables of micro objects $\Delta x$ and $%
\Delta p$ beyond the Heisenberg inequality $\Delta x\cdot \Delta p\geq \hbar
.$ Strong fields are adequately enough described by the classical gravity
theory, so we do not consider these in the present study. Let us emphasize
that the assumption on existence of such a negligibly small background is
quite natural. With this, we assume the gravity background to be
isotropically distributed over the space.

There exist certain technical difficulties now in implementation of a
quantum computer. So let us construct a quasi-classical model of such a
quantum computer scheme, employing the valid classical radio-engineering
model [3] of the ESR experiment. It is provided by sending two obzervers
electromagnetic pulses with the same stochastic phase. In this model of the
quantum computer, employed are continuous excitation voltage, parametric
generators in the number equal to the number of qubits, digital time-delay
lines, multipliers of logical signals and low-pass filters. Provided here is
two-validness of the function $sign\{x\}$, as quantum-mechanical spin. To
implement the described procedure, let us modulate the phase of the carrier
monochromatic oscillation $X(t)$ with the frequency $\omega _{0}$ by a
random process $\phi (t)$, $X(t)=\cos [\omega _{0}t+\phi (t)]$ (we assume
the correlation time for the process $\phi (t)$ to be much higher than the
period of oscillations). Further, we introduce into the oscillation $X(t)$
the controlled phase shift $\alpha $ and mix it with the ''homodyne''
oscillation $\cos (\omega _{0}t)$ possessing the stable phase. The obtained
superposition $Z(t)=\cos [\omega _{0}t+\phi (t)+\alpha ]+\cos (\omega _{0}t)$
is then rectified. At the output of the square-law detector \{after having
filtered out the high-frequency component with the frequency of $2\omega _{0}
$, we shall get the low-frequency signal $Z(t)\thickapprox 2+2\cos [\phi
(t)+\alpha ]$. This results in $\lambda $ correlated cells modulated with
the same random process $\phi (t)$, which in our case simulates the hidden
variables.

In conclusion, such a quantum computer will be efficient for a special class
of problems; with this, it would require the respective software.

\end{document}